\documentstyle [12pt,epsf]{article}

\input{epsf}
\begin{document}
\begin{titlepage}
\begin{center}
\vspace{0.4in} {\Large \bf
Scalar Field Theory  at Finite Temperature in D=2+1}\\
\vspace{.3in} {\large\em Gino N.~J.~A\~na\~nos
\\
 Instituto de F\'{\i}sica Te\'orica-IFT\\
 Universidade Estadual Paulista
   \\ Rua Pamplona
145, S\~ao Paulo, SP  01405-900  Brazil
 \\
 E-mail: gananos@ift.unesp.br }

\subsection*{Abstract}
\end{center}
We discuss the  $\varphi^6 $ theory defined in $D=2+1$-dimensional
space-time and assume that the system is in equilibrium with a
thermal bath at temperature $\beta^{-1}$. We use the  $ 1/N $
expansion and the method of the composite operator (CJT) for
summing a large set of Feynman graphs.We demonstrate explicitly
the Coleman-Mermin-Wagner theorem at finite temperature.


\end{titlepage}

\newpage

\baselineskip .37in
\section {Introduction}

The conventional perturbation theory in the coupling constant or in
$\hbar$, i.e., the loop expansion can only be used for the study of
small quantum corrections to classical results. When discussing
quantum mechanical effects to any given order in such an expansion,
one is not usually able to justify the neglect of yet higher
 order. In other words, for theories with a large $N$ dimensional internal
symmetry group, there exist another perturbation scheme, the $1/N$ expansion,
which circumvents this criticism. Each term in the $ 1/N $ expansion
contains an infinite subset of terms of the loop expansion.
The $ 1/N $ expansion has the nice property that the leading-order quantum
 corrections are of the same order as the classical quantities. Consequently, the leading
 order which adequately characterizes the theory in the large $N$ limit preserves much of
 the nonlinear structure of the full theory.

The scalar field  $(\varphi^4)_{D=4}$ theory at finite temperature is
of great interest in the field of phase transitions in the early
universe and heavy ion collisions. When used as a simple model for
the Higgs particle in the standard model of electroweak
interactions, it may allow the study of symmetry breaking phase
transitions in the early universe. For $N=4$ scalar fields, it is
also a model of chiral symmetry breaking in QCD and hence is
relevant for the theoretical study of heavy ion collisions.
Moreover, this theory is an excellent theoretical laboratory in
which analytic nonperturbative methods can be tested. Now for the
case $D=2+1$ it has been shown that, in the large $N$ limit, the
$\varphi^6$ theory has a UV fixed point and therefore must have a
second IR fixed point~\cite{bardeen} and for this we could say that
at least for large $N$ the $(\varphi^6)_{D=3}$ theory is known to be
qualitatively different from $(\varphi^4)_{D=4}$ theory. For other
ways, theories in less than four space-time dimensions can offer
interesting and complex behavior as well as tractability, and for
example the case of three space-time dimensions, they can even be
directly physical, describing various planar condensed matter
systems. For example the introduction of the $\varphi^6$ term
generates a rich phase diagram, with the possibility of second
order, first order phase transitions or even both transitions
occurring simultaneously. This situation defines the tricritical
phenomenon. For example some systems such antiferromagnets in the
presence of a strong external field or the $He^3$-$He^4$ mixture
exhibits such behavior.

\section{The effective potential}
The theory for which we are interested is given by the Lagrangian,
\begin{equation}
{\cal L}(\varphi)=\frac{1}{2} (\partial_{\mu}
\varphi)^2-\frac{1}{2}m_{0}^2 \varphi^2-\frac{\lambda_0}{4!N}
\varphi^4- \frac{\eta_0}{6! N^2}\varphi^6,
\end{equation}
is a theory of $N$ real scalar fields with $O(N)$ symmetry.

For definiteness, we work at zero temperature; however, the finite
temperature generalizations can be easily obtained \cite{dolan}.

In this section we are going to use the method of composite
operator developed by Cornwall, Jackiw and Tomboulis (CJT)
~\cite{cornwall,livro} in order to get the effective potential
$\Gamma (\phi)$ at leading order in the $1/N$ expansion. The
composite operator formalism reduces the problem to summing two
particle irreducible (2PI) Feynman graphs by defining a
generalized effective action $\Gamma (\phi,G)$ which is a
functional not only of $\phi_a (x)$, but also of the expectation
values $G_{ab} (x,y)$ of the time ordered product of quantum
fields $<0|T(\varphi(x)\varphi(y))|0> $, i.e.
\begin{equation}
 \Gamma (\phi,G)= I(\phi)+\frac{i}{2} Tr\,Ln\, G^{-1} +
\frac{i}{2} Tr\, D^{-1}(\phi) G + \Gamma_2(\phi,G) +\dots \;\; ,
\label{G1}
\end{equation}
where $I(\phi)=\int dx^D \,{\cal L}(\phi) $, $G$ and $D$ are
matrices in both the functional and the internal space whose
elements are $G_{ab}(x,y)$, $D_{ab}(\phi;x,y)$, respectively and
$D$ is defined by
\begin{equation}
i D^{-1}=\frac{\delta^2 I(\phi)}{\delta\phi(x) \delta\phi(y)}.
 \end{equation}
The quantity $\Gamma_2(\phi,G)$ is computed as follows. In the
classical action $I(\varphi)$ we must shift the field $\varphi$
by $\phi$. The new action  $I(\varphi+\phi)$  possesses terms
cubic and higher in $\varphi$. This defines an interaction part
$I_{int} (\varphi,\phi)$ where the vertices depend on $\phi$.
$\Gamma_2(\phi,G)$ is given by sum of all  (2PI) vacuum graphs  in
a theory with vertices determined by $I_{int} (\varphi,\phi)$ and
the propagators set equal to $G(x,y)$. The trace and logarithm in
Eq.(\ref{G1}) are functional. After these procedures the
interaction Lagrangian density becomes
\begin{eqnarray}
 {\cal L}_{int}(\varphi,\phi)&=&-\frac{1}{2} \left( \frac{\lambda_0 \phi_a}{3N} + \frac{\eta_0 \phi^2\phi_a}{30N^2} \right) \varphi_a \varphi^2 -
 \left( \frac{8\eta_0 \phi_a\phi_b\phi_c}{6N^2} \right) \varphi_a \varphi_b
\varphi_c -\frac{1}{4!N} \left( \lambda_0+ \frac{\eta_0\phi^2}{10N}\right)
 \varphi^4 \nonumber \\
& &
-\left( \frac{12\eta_0\phi_a\phi_b}{6!N^2} \right)\varphi_a\varphi_b\varphi^2
-\frac{1}{5!}\left( \frac{\eta_0\phi_a}{N^2}\varphi_a\varphi^4\right)
-\frac{\eta_0}{6!N^2}\varphi^6.
\end{eqnarray}
The effective action $\Gamma(\phi)$ is found by solving for $G_{ab}(x,y)$  the equation
\begin{equation}
\frac{\delta \Gamma(\phi,G)}{\delta G_{ab}(x,y)}=0,
\label{eq1}
\end{equation}
and substituting the solution in the generalized effective action
$\Gamma(\phi,G)$.
The vertices in the above equation contain
factors of $1/N$ or $1/N^2$, but a $\varphi$ loop gives a factor
of $N$ provided the $O(N)$ isospin flows around it alone and not
into another part of the graph. We usually call such loops
bubbles. Then at leading order in $1/N$, the vacuum graphs are
bubble trees with two or three bubbles at each vertex. The (2PI)
graphs are shown in Fig. 1.
\begin{figure}[ht]

\centerline{\epsfysize=1.0in\epsffile{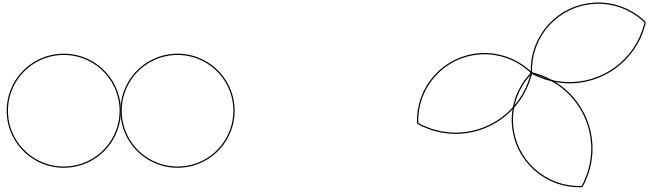}}
\caption[region]
{The 2PI vacuum graphs.}
\end{figure}
It is straightforward to obtain
\begin{equation}
\Gamma_2(\phi,G)=\frac{-1}{4!N}\int d^{D}x \left( \lambda_0+
\frac{\eta_0\phi^2}{10N} \right) [G_{aa}(x,x)]^2 -
\frac{\eta_0}{6!N^2}
\int d^{D}x [G_{aa}(x,x)]^3 .
\label{eq2}
\end{equation}

Therefore Eq.(\ref{eq1}) becomes

\begin{eqnarray}
\frac{\delta\Gamma(\phi,G)}{\delta G_{ab}(x,y)}&=& \frac{1}{2}(G^{-1})_{ab}
(x,y) + \frac{i}{2} D^{-1}(\phi)-\frac{1}{12N}\left( \lambda_0+
\frac{\eta_0\phi^2}{10N}\right)[\delta_{ab}G_{cc}(x,x)]\delta^D(x-y)
\nonumber \\
& & -\frac{3\eta_0}{6!N} \delta_{ab}[G_{cc}(x,x)]^2\delta^D(x-y)=0.
\label{eq3}
\end{eqnarray}
Rewriting this equation, we obtain the gap equation
 \begin{eqnarray}
(G^{-1})_{ab}(x,y) &=& D^{-1}_{ab}(\phi;x,y)+ \frac{i }{6N}
\left( \lambda_0+
\frac{\eta_0\phi^2}{10N}\right)[\delta_{ab}G_{cc}(x,x)]\delta^D(x-y)
\nonumber
\\
 & &+\frac{i\eta_0}{5!N^2} \delta_{ab}[G_{cc}(x,x)]^2\delta^D(x-y).
\label{eq4}
\end{eqnarray}
Hence
\begin{equation}
\frac{i}{2} Tr\, D^{-1}G= \frac{1}{12N}\int d^Dx
\left( \lambda_0+\frac{\eta_0\phi^2}{10N} \right) [G_{aa}(x,x)]^2 +
\frac{3\eta_0}{6!N^2}\int d^{D}x [G_{aa}(x,x)]^3+ cte .
\label{eq5}
\end{equation}
Using Eqs.(\ref{eq4}) and (\ref{eq5}) in Eq.(\ref{eq2}) we find the effective action
\begin{eqnarray}
\Gamma(\phi) &=& I(\phi)+\frac{i}{2} Tr\,[Ln\, G^{-1}]+\frac{1}{4!N}
\int d^Dx \left( \lambda_0+\frac{\eta_0\phi^2}{10N} \right) [G_{aa}(x,x)]^2
 \nonumber \\
& & +\frac{2\eta_0}{6!N^2} \int d^Dx [G_{aa}(x,x)]^3 ,
\label{effa}
\end{eqnarray}
where $G_{ab}$ is given implicitly by Eq.(\ref{eq4}). The trace in
(\ref{effa}) are both the functional and the internal space.
The last two terms on the right hand side of Eq.(\ref{effa}) are the leading
contribution to the effective action in the $1/N$ expansion.
As usual we may simplify the situation by separating $G_{ab}$ into
transverse and longitudinal components, so
\begin{equation}
G_{ab}=\left(\delta_{ab}-\frac{\phi_a \phi_b}{\phi^2}\right)g+
\frac{\phi_a \phi_b}{\phi^2} \stackrel{\sim}{g}\; ,
\end{equation}
in this form we can invert $G_{ab}$,
\begin{equation}
(G)^{-1}_{ab}=\left(\delta_{ab}-\frac{\phi_a \phi_b }{\phi^2} \right) g^{-1} +
\frac{\phi_a \phi_b}{\phi^2} {\stackrel{\sim}{g}}^{-1}\; .
\end{equation}

Now we can take the trace with respect to the indices of the internal
space,
\begin{equation}
G_{aa}=Ng +O(1) ,\;\;\;\; (G)^{-1}_{aa}=Ng^{-1}+ O(1) \; .
\label{gd}
\end{equation}
From this equation at leading order in $1/N$, $G_{ab}$ is diagonal
in $a,b$. Substituting Eq.(\ref{gd}) into Eq.(\ref{effa}) and
Eq.(\ref{eq4}) and keeping only the leading order one finds that
the daisy and superdaisy resummed effective potential for the
$\varphi^6$ theory is given by:
\begin{eqnarray}
\Gamma(\phi)&=&I(\phi)+\frac{iN}{2} tr\,(\ln\, g^{-1})+\frac{N}{4!}
\int d^Dx \left( \lambda_0+\frac{\eta_0\phi^2}{10N} \right) g^2(x,x)
\nonumber \\
 & & +\frac{2N\eta_0}{6!} \int d^Dx g^3(x,x) +O(1) ,
\label{effa1}
\end{eqnarray}
where the trace is only in the functional space, and the gap equation
becomes
\begin{equation}
g^{-1}(x,y)=i\left[ \Box + m_0^2+
\frac{\lambda_0}{6}\left(\frac{\phi^2}{N}+ g(x,x)\right) + \frac{\eta_0}{5!}
\left(\frac{\phi^2}{N}+ g(x,x)\right)^2 \right ] \delta^{D}(x-y)+ O\left(\frac{1}{N}\right)
\label{gap1}.
\end{equation}
It is important to point out that this calculation was done by
Townsend \cite{townsend}.
\section{The theory at finite temperature}
In order to generalize these results to the case of finite
temperature we are going to assume that the system is in equilibrium
with a thermal bath a temperature $T=\beta^{-1}$. Since we are
interested in the equilibrium situation it is convenient to use the
Matsubara formalism (imaginary time). Consequently it is convenient
to continue all momenta to Euclidean values $(p_0=ip_4)$ and take
the following Ansatz for $g(x,y)$:
\begin{equation}
g(x,y)=\int
\frac{d^{D}p}{(2\pi)^D}\frac{\exp^{i(x-y)p}}{p^2+M^2(\phi)}.
\label{ge}
\end{equation}
Substituting Eq.(\ref{ge}) in Eq.(\ref{gap1}) we get the
expression for the gap equation,
\begin{equation}
 M^2(\phi)=m^2_0+
\frac{\lambda_0}{6}\left( \frac{\phi^2}{N}+F(\phi)\right) +
\frac{\eta_0}{5!} \left( \frac{\phi^2}{N}+ F(\phi)\right)^2,
\label{M2}
\end{equation}
where $F(\phi)$ is given by
\begin{equation}
F(\phi)=\int \frac{d^Dp}{(2\pi)^D}\frac{1}{p^2+M^2(\phi)},
\label{Fphi}
\end{equation}
and the effective potential in the $D$-dimensional Euclidean space
can be written as
\begin{equation}
V(\phi)=V_0(\phi)+\frac{N}{2}\int \frac{d^Dp}{(2\pi)^D} \ln \left
[p^2+M^2(\phi) \right ]
 -\frac{N}{4!}\left(\lambda_0+\frac{\eta_0 \phi^2}{10N}\right) F(\phi)^2-
\frac{2N\eta_0 F(\phi)^3}{6!} , \label{ve1}
\end{equation}
where $V_0(\phi)$ is the tree approximation of the potential.

Replacing the continuous four momenta $k_4$ by discrete $\omega_n$
and the integration by a summation ($\beta=1/T$). The
effective potential at finite temperature is
\begin{eqnarray}
V_{\beta}(\phi)&= &
V_0(\phi)+\frac{N}{2\beta}\sum_{n}^{\infty} \int
\frac{d^{D-1}p}{(2\pi)^{D-1}} \ln \left
[\omega_n+p^2+M_{\beta}^2(\phi) \right ]
\nonumber \\
& &-\frac{N}{4!}\left(\lambda_0+\frac{\eta_0 \phi^2}{10N}\right)
F_{\beta}(\phi)^2-\frac{2N\eta F_{\beta}(\phi)^3}{6!},
\label{ve3}
\end{eqnarray}
where the expression $F_{\beta}(\phi)$ is the finite temperature
generalization of $F(\phi)$, and is given by
\begin{equation}
F_{\beta}(\phi)=\frac{1}{\beta}\sum_{n=-\infty}^{\infty} \int
\frac{d^{D-1}p}{(2\pi)^{D-1}}
\frac{1}{\omega^2_n+p^2+M^2_{\beta}(\phi)}\; . \label{Fbphi}
\end{equation}
The gap equation for this theory at finite temperature is
\begin{equation}
 M^2_{\beta}(\phi)=m^2_0+
\frac{\lambda_0}{6}\left( \frac{\phi^2}{N}+F_{\beta}(\phi)\right)
+ \frac{\eta_0}{5!} \left( \frac{\phi^2}{N}+
F_{\beta}(\phi)\right)^2. \label{M2T1}
\end{equation}
 In order to regularize
$F_{\beta}(\phi)$ given by Eq.(\ref{Fbphi}), we use a mixing
between dimensional regularization and analytic regularization.
For this purpose we define the following expression,
 \begin{equation}
I_{\beta}(D,s,m)=\frac{1}{\beta}\sum_{n=-\infty}^{\infty}\int
\frac{d^{D-1}k}{(2\pi)^{D-1}}\frac{1}{(\omega^2_n+k^2+m^2)^s} \; .
\end{equation}
The analytic extension of the inhomogeneous Epstein zeta function
can be done and the corresponding  analytic extension of
$I_{\beta}(D,s,m)$ is
\begin{equation}
I_{\beta}(D,s,m)= \frac{m^{D-2s}}{(2\pi^{1/2})^D\Gamma(s)} \left[
\Gamma\left(s-\frac{D}{2}\right) + 4 \sum_{n=1}^{\infty} \left(
\frac{2}{mn\beta}
 \right)^{D/2-s} K_{D/2-s}(mn\beta) \right]
\end{equation}
where $K_{\mu}(z)$ is the modified  Bessel function of the third
kind. Fortunately for $D=2+1$ the analytic extension of the
function $I_{\beta}(D,s=1,m=M_\beta(\phi))=F_{\beta}(\phi)$ is
finite and can be expressed in a closed form~\cite{an}
\begin{equation}
F_{\beta}(\phi)=I_{\beta}(3,1,M_{\beta}(\phi))=
-\frac{M_{\beta}(\phi)}{4\pi}\left(
1+\frac{2\,\ln(1-e^{-M_{\beta}(\phi)\beta})}{M_{\beta}(\phi)\beta}
\right ). \label{m2b2}
\end{equation}
We note that in $D=2+1$ we have no pole, at least in this
approximation. To proceed to regularize the second term of
Eq.(\ref{ve3}), we define
\begin{equation}
LF_\beta(\phi)=\frac{1}{\beta}\sum_{n=1}^{\infty} \int
\frac{d^{D-1}p}{(2\pi)^{D-1}} \ln \left
[\omega_n+p^2+M_{\beta}^2(\phi) \right ]
\end{equation}
then,

 \begin{equation}
\frac{ \partial LF_\beta(\phi)}{\partial
M_{\beta}}=(2M_{\beta})\frac{1}{\beta}\sum_{n=1}^{\infty} \int
\frac{d^{D-1}p}{(2\pi)^{D-1}}
\frac{1}{\omega_n+p^2+M_{\beta}^2(\phi)}
\end{equation}
and from Eq.(\ref{Fbphi}), we have that
 \begin{equation}\label{lfb}
\frac{ \partial LF_\beta(\phi)}{\partial M_{\beta}}=(2M_{\beta})
F_{\beta}(\phi).
\end{equation}
For  $D=2+1$, $F_{\beta}(\phi)$ is finite and is given by
Eq.(\ref{m2b2})~\cite{an} and integrating the Eq.(\ref{lfb}), we
obtain
\begin{equation}\label{m2b2}
LF_\beta(\phi)_R=-\,{\frac {{M_{\beta}(\phi)}^{3}}{6\pi }}-{\frac
{M_{\beta}(\phi) Li_2({e^{-M_{\beta}(\phi)\beta}})} {\pi
\,{\beta}^{2}}}-{\frac {Li_3({e^{-M_{\beta}(\phi)\beta}})}{\pi
\,{\beta}^{3}}} .
\end{equation}
The definition of general polylogarithm function $Li_n(z)$ can be found in
Ref.~\cite{poly}.

The daisy and super daisy resummed effective potential at finite
temperature for $D=2+1$ is then given by
\begin{equation}
V_{\beta}(\phi)=V_0(\phi)+\frac{N}{2} LF_\beta(\phi)_R-
\frac{N}{4!}\left(\lambda_0+\frac{\eta_0 \phi^2}{10N}\right)
(F_{\beta}(\phi)_R)^2 -\frac{2N\eta (F_{\beta}(\phi)_R)^3}{6!}
\label{ve4}
\end{equation}
and the corresponding gap equation [see Eq.(\ref{M2T1})]:
\begin{eqnarray}\label{m2d3}
 M^2_{\beta}(\phi) & = &  m^2_0+
\frac{\lambda_0}{6}\left(
\frac{\phi^2}{N}-\frac{M_{\beta}(\phi)}{4\pi} \left[
1+\frac{2\,\ln(1-e^{-M_{\beta}(\phi)\beta})}{M_{\beta}(\phi)\beta}
\right ]\right)  \nonumber \\
&& + \frac{\eta_0}{5!} \left(
\frac{\phi^2}{N}-\frac{M_{\beta}(\phi)}{4\pi} \left[
1+\frac{2\,\ln(1-e^{-M_{\beta}(\phi)\beta})}{M_{\beta}(\phi)\beta}
\right]\right)^2 .
\end{eqnarray}
From this expression we can deduce that there is no possible way to
find a solution for $M_\beta$ going to zero, because the terms
in Eq.(\ref{m2d3}) containing the logarithm will not permit, and this
situation is similar to the scalar theory with $O(N)$ symmetry in
$2D$ at zero temperature~\cite{coleman1}. This result is in
agreement with the the Coleman-Mermin-Wagner theorem
~\cite{coleman2}, which statement is related to the fact that it is
impossible to construct a consistent theory of massless scalar in
$2D$. If a spontaneous breaking of continuous symmetry were to
happen at finite $T$, then one would be faced with this problem at
momentum scales below $T_c$, i.e. it would be impossible to
construct an effective $2D$ theory of the Goldstone bosons
zero modes.

\section{Conclusions}
In this paper we have found the daisy and super daisy effective
potential for the theory $\varphi^{6}$ in $D=2+1$-dimensional
Euclidean space at finite temperature. The form of effective
potential have been found explicitly using resummation method in
the leading order $1/N$ approximation (Hartree-Fock
approximation). We found that in this approximation there is no
symmetry breaking for any temperature and this is clearly a
manifestation of the Coleman-Mermin-Wagner theorem which
stipulates that the spontaneous symmetry breaking of continuous
symmetry cannot happen ind $D=2+1$ at finite temperature.
\section{Acknowlegement}

This paper was supported by FAPESP under contract number 03/12271-7.

\end{document}